\documentclass[12pt,preprint]{aastex}

\slugcomment{To appear in $The$ $Astrophysical$ $Journal$}

\shorttitle{Elements in the s- and r-Process-rich Stars}
\shortauthors{Zhang, Ma, & Zhou}

\begin{document}
\title {Neutron-capture elements in the s- and r-process-rich stars: Constraints on neutron-capture nucleosynthesis processes}
\author{Bo Zhang, Kun Ma, and Guide Zhou}
\affil{Department of Physics, Hebei Normal University, Shijiazhuang 050016, China;\\
zhangbo@hebtu.edu.cn}

\begin{abstract}
The chemical abundances of the very metal-poor double-enhanced
stars are excellent information for setting new constraints on
models of neutron-capture processes at low metallicity. These
stars are known as \emph{s}+\emph{r} stars, since they show
enhancements of both s-process and r-process elements. The
observed abundance ratios for the double-enhanced stars can be
explained by those of stars that were polluted by an AGB star and
subsequently accreted very significant amounts of r-process
material out of an AIC (accretion-induced collapse) or Type 1.5
supernova. In this paper we present for the first time an attempt
to fit the elemental abundances observed in the s- and r-rich,
very metal-poor stars using a parametric model and suggest a new
concept of component coefficients to describe the contributions of
the individual neutron-capture processes to double-enhanced stars.
We find that the abundance ratios of these stars are best fitted
by enrichments of s- and r-process material. The overlap factor in
the AGB stars where the observed s-process elements were produced
lies between 0.1 and 0.81. Taking into account the dependence of
the initial-final mass relations on metallicity, this wide range
of values could possibly be explained by a wide range of core-mass
values of AGB stars at low metallicity. The component coefficient
of the r-process is strongly correlated with the component
coefficient of the s-process for the double-enhanced stars. This
is significant evidence that the r-process material in
double-enhanced stars comes from an AIC or Type 1.5 supernova.

\end{abstract}


\section{Introduction}
The elements heavier than those belonging to the iron peak are
mostly made through neutron capture via two principal processes:
the r-process (for rapid neutron-capture process) and the
s-process \citep[for slow neutron-capture process;][]{bur57}.
During quiescent He burning, neutrons generated via the
$^{13}$C($\alpha$, \emph{n})$^{16}$O \citep{cam55,ree66,mat85} and
$^{22}$Ne($\alpha$, \emph{n})$^{25}$Mg \citep{cam60} reactions are
captured by heavy-element seed nuclei. The neutron densities are
relatively low ($N_n\sim10^8 cm^{-3}$), so nearly all possible
$\beta$-decays will have time to occur between successive neutron
captures. Successful synthesis of heavier isotopes progresses
along the ``valley of $\beta$ stability." This synthesis route is
called the s-process, and it is responsible for the production of
about half of the isotopes of the \emph{n}-capture elements. The
overabundances of elements heavier than iron observed at the
surface of asymptotic giant branch (AGB) stars \citep{smi90}
clearly indicate that the s-process takes place during the AGB
phase in the evolution of low- and intermediate-mass stars
($0.8\leq M/M_\odot\leq 8$). The site or sites of the r-process
are not known, although suggestions include the $\nu$-driven wind
of Type II supernovae \citep{woo92,woo94}, the mergers of neutron
stars \citep{lat74,ros00}, AIC \citep[accretion-induced
collapse;][]{qia03}, and Type 1.5 supernovae \citep{zij04}.
Extremely neutron rich nuclei are formed in a matter of seconds in
the r-process, and it is not necessary to have preexisting
heavy-element seed nuclei. The r-process is also responsible for
about half of the abundances of solar system n-capture isotopes.
Thus, n-capture elements can be composed of some pure r-process,
pure s-process, and some mixed-parentage isotopes. As a result,
when the solar system total abundances ($t_{ss}$) are separated
into contributions from the s-process ($s_{ss}$) and the r-process
($r_{ss}$), some elements are mostly contributed by the r-process,
such as Eu, and some by the s-process, such as Ba. Therefore, Eu
is commonly referred to as an ``r-process element,'' and Ba, as an
``s-process element.''

Most metal-poor stars in the Galaxy show that about 14\%
 of those are carbon enhanced
\citep[hereafter CEMP stars;][]{coh05}. These stars are often
enriched in neutron-capture heavy elements. However, the
enhancement of s-process and r-process elements varies from star
to star.\ Observations of metal-poor stars with metallicities
lower than [Fe/H]$\approx-2.5$ enriched in neutron-capture
elements have revealed the solar r-process pattern, while only two
cases of  highly r-process-enhanced stars (hereafter ``\emph{r}"
stars), CS 22892-052 \citep{sne96,sne03} and CS 31082-001
\citep{cay01,hil02}, have been noted. Despite their considerable
metal deficiency, these stars seem to have experienced an
r-process that barely differs from the sum of r-processes that
enriched the pre-solar nebula. This has led to suggestions that
r-process production may be independent of the initial metallicity
of the site, especially for the heavier n-capture elements
\citep[Z$\geq 56$;][]{cow95,sne96,nor97,sne00}.

Although the material from which Population II stars form is not
expected to contain significant s-process contributions, some
stars including some subgiants are greatly enriched in carbon and
s-elements \citep[hereafter \emph{s} stars;][]{nor97,hil00}.\
These are believed to be binary companions of initially more
massive donor stars that have evolved through the thermally
pulsing AGB phase and transferred material enriched in C and
s-process elements onto the lower mass, longer lived secondary now
observed.

The discovery that several stars show enhancements of both
r-process and s-process elements \citep{hil00,coh03} is puzzling,
as they require pollution from both an AGB star and a supernova.
\citet{qia03} proposed a theory for the creation of
``\emph{s}+\emph{r}"-process stars. First some s-process material
is accreted from an AGB star, which turns into a white dwarf.
Later in the evolution of the system, the white dwarf accretes
matter from the polluted star and suffers an AIC to a neutron
star. The $\nu$-driven wind produces an r-process, which also
pollutes the companion. A possible problem these authors mention
is the still uncertain nucleosynthesis in accretion induced
collapse, which may or may not produce the r-process. Another
possible \emph{s}+\emph{r} scenario is that the AGB star transfers
s-rich matter to the observed star but does not suffer a large
mass loss and at the end of the AGB phase; the degenerate core of
the low-metallicity, high-mass AGB star may reach the
Chandrasekhar mass, leading to a Type-1.5 supernova \citep{zij04}.
Such supernovae can explain both the enhancement pattern and the
metallicity dependence of the double-enhanced halo stars.

There is another possibility for the double-enhanced scenario of
the halo stars. In this picture, the formation of a binary system
of low-mass stars was triggered by a supernova that polluted and
clumped a nearby molecular cloud. Subsequently, the observed star,
already strongly enhanced in r-process elements, receives large
amounts of s-process elements from the initially more massive star
that underwent the AGB phase, and turns into the double-enhanced
star \citep[hereafter``r+s" stars;][]{del04,bar05,gal05,iva05}.
\citet{aoki02} proposed that this scenario could possibly explain
the formation of the double-enhanced stars, but the r-rich very
metal-poor stars are so rare that it is implausible to explain the
larger number of double-enhanced stars through even rarer ones.
Previously, \citet{bar05} and \citet{wan06} suggested massive AGB
stars (M $\approx$ 8$-$12M$_\odot$) to be the origin of these
double enhancements. Such a large-mass AGB star could possibly
provide the observed enhancement of s-process elements in the
first phase and explode or collapse providing the r-process
elements, but the modeling of the evolution of such a large-mass
metal-poor star is a difficult task, and the amount of the
s-process material produced and its abundance distribution is
still uncertain \citep{wan06}.

It is now widely accepted that the neutron exposure required to
produce s-elements in AGB stars originates in some partial mixing
of protons (PMP) from the envelope down into the C-rich layers
resulting from the former intermittent operation of He burning
during the thermal pulse phase of AGB stars. PMP activates the
chain of reactions
$^{12}$C(\emph{p},$\gamma$)$^{13}$N($\beta$)$^{13}$C($\alpha$,\emph{n})$^{16}$O.
The s-elements thus produced in the deep interior by successive
neutron captures are subsequently brought to the surface by the
third dredge-up \citep{gal98}. Using the primary-like neutron
source [$^{13}$C($\alpha$,\emph{ n})$^{16}$O] and starting with a
very low initial metallicity, most iron seeds are converted into
$^{208}$Pb. Thus, when the third dredge-up episodes mix the
neutron-capture products into the envelope, the star appears
s-enhanced and lead-rich. Therefore, if the standard PMP scenario
holds, all s-process-enriched AGB stars with metallicities
[Fe/H]$\leq$-1.3 are thus predicted to be Pb stars
([Pb/hs]$\geq$1, where hs denotes the `heavy' s-process elements
such as Ba, La, and Ce), independent of their mass and metallicity
\citep{gor00}.

The first three ``lead" stars (HD 187861, HD 224959, and HD
196944) have been reported by  \citet{eck01}. At the same time,
\citet{aoki01} found that the slightly more metal-deficient stars
LP 625-44 and LP 706-7 are enriched in s-elements but cannot be
considered as lead stars ([Pb/Ce]$\leq$0.4), in disagreement with
the standard PMP predictions. A large spread of $^{13}$C pocket
efficiencies is proposed by \citet{str05} in order to explain the
spreads of [Pb/\emph{hs}]. However, a physical explanation for the
different $^{13}$C pocket strengths has not been yet found, so
fundamental uncertainties currently exist in the models of AGB
stars.

The nucleosynthesis of neutron-capture elements in CEMP stars can
be investigated by abundance studies of s-rich and r-rich stars.
Recently, \citet{coh03}, by analysing the spectra of the
double-rich star HE 2148-1247, concluded that the observed
abundances could not be fitted by a scaled  solar system
s-process, r-process, and total solar inventory. A quantitative
understanding of the origins of neutron-capture elements in the
double-enhanced halo stars has so far been a challenging problem.
Although some of the basic tools for this task were presented
several years ago, the origins of the neutron-capture elements in
the double-enhanced halo stars, especially r-process elements, are
not clear, and the characteristics of the s-process
nucleosynthesis in the AGB stars are not ascertained. Clearly, the
study of elemental abundances in these objects is important for
investigation of the origin of neutron-capture elements in the
objects and in the Galaxy.

In this paper we investigate the characteristics of the
nucleosynthesis pathway that produces the abundance ratios of
double-rich objects using the s-process parametric model and
considering very significant amounts of enrichment of r-process
material out of an AIC or Type 1.5 supernova(SN 1.5). The
parametric model of double-enhanced stars is described in Sect.\
2. The calculated results are presented in Sect.\ 3, in which we
also discuss the characteristics of the s- and r-process
nucleosynthesis. Conclusions are given in Sect.\ 4.

\section{Parametric Model of Double-enhanced Stars}

There have been many theoretical studies of s-process
nucleosynthesis in low-mass AGB stars.\ Unfortunately, however,
the precise mechanism for chemical mixing of protons from the
hydrogen-rich envelope into the $^{12}$C-rich layer to form a
$^{13}$C pocket is still unknown \citep{bus01}. This makes it even
harder to understand the particular abundance pattern of the s-
and r-process elements found in carbon-rich, metal-poor stars. It
is interesting to adopt the parametric model for metal-poor stars
presented  by \citet{aoki01}, with many of the neutron-capture
rates updated \citep{bao00}, to study which physical conditions
reproduce the observed abundance patterns found in the stars. On
one hand, this approach has the virtue of being model independent,
in that it does not refer to detailed stellar evolution models. On
the other hand, strictly speaking, our model does not represent
exactly the situation proposed by current stellar s-process
models. In the parametric model, in fact, we make use of an
exponential distribution of neutron exposures via sequential
irradiations \citep{how86}. This representation would well
describe a case in which the $^{13}$C neutron source burns during
thermal pulses. In the current stellar s-process model, instead,
the $^{13}$C neutron source is activated in radiative conditions
in the periods between thermal pulses \citep{str95}, and its
activation cannot be approximated by a simple exponential law of
neutron irradiations \citep{gal98}. There are four parameters in
the parametric model on s-process nucleosynthesis. They are the
neutron irradiation time $\Delta$t, the neutron number density
$N_n$, the temperature
 $T_9$ (in units of 10$^9$ K) at the onset of the s-process, and the
overlap factor \emph{r}. Combining these quantities, we can obtain
the neutron exposure per thermal pulse,
$\Delta\tau=N_{n}v_{T}\Delta t$, where $v_T$ is the average
thermal velocity of neutrons at $T_9$. The temperature and the
neutron number density $N_n$ are fixed respectively at reasonable
values for the $^{13}$C($\alpha$,n)$^{16}$O reaction, $T_9=0.1$ K
and $N_n$=10$^7$ cm$^{-3}$, for these studies. Thus, the final
s-process abundance distributions depend only on the neutron
exposure $\Delta\tau$ and overlap factor \citep{aoki01}.

We explored the origin of the neutron-capture elements in the
double-enhanced stars by comparing the observed abundances with
predicted r- and s-process contributions. For this purpose, we
propose that the \textit{i}th element abundance in the star can be
calculated as \begin{equation}
N_{i}(Z)=C_{s}N_{i,\ s}+C_rN_{i,\ r}10^{[Fe/H]} ,
\end{equation}
where \emph{Z} is the metallicity of the star, $N_{i,\ s}$ is the
abundance of the \textit{i}th element produced by the s-process in
the AGB star and $N_{i,\ r}$ is the abundance of the \textit{i}th
element produced by the r-process (per Si=$10^6$ at Z=Z$_\odot$),
and $C_s$ and $C_r$ are the component coefficients that correspond
to contributions from the s-process and the r-process,
respectively. It should be noted that the s-process abundances in
the envelopes of the stars could be expected to be lower than the
abundance produced by the s-process in the AGB star, because the
material is mixed with the envelopes of the primary (former AGB
star) and secondary stars.

Based on equation (1), we carry out the s-process nucleosynthesis
calculation combined with the contribution of the r-process to fit
the abundance profile observed in the double-enhanced stars, in
order to look for the minimum $\chi^2$ in the four-parameter space
formed by \emph{r}, $\Delta\tau$, $C_s$, and $C_r$. The adopted
initial abundances of seed nuclei lighter than the iron peak
elements were taken to be the solar-system abundances, scaled to
the value of [Fe/H] of the star. Because the
neutron-capture-element component of the interstellar gas that
formed very metal-deficient stars is expected to consist of mostly
pure r-process elements, for the other heavier nuclei we use the
solar system r-process abundances \citep{arl99}, normalized to the
value of [Fe/H].

The ultra-metal-poor star CS 22892-052 merits special attention,
because this star has an extremely large overabundance of
n-capture elements relative to iron and very low metallicity with
[Fe/H]$=-3.1$. Many studies
\citep{cow99,sne96,sne98,sne00,nor97,pfe97} have suggested that
the abundance patterns of the heavier (Z$\geq$56) stable
neutron-capture elements in very metal-poor stars are consistent
with the solar system r-process abundance distribution. This
concordance breaks down for the lighter neutron-capture elements
in the range of 40$<$Z$<$56 \citep{sne00}. \citet{zha02} have
reported that when the abundances of the lighter elements in CS
22892-052 are multiplied by a factor of 1/0.4266, the abundance
distributions obtained for both heavier and lighter
neutron-capture elements are in accordance with the solar system
r-process pattern. This star could well have abundances that
reflect the nucleosynthesis of  a single supernova \citep{fie02},
so the adopted abundances of nuclei $N_{i,\ r}$ in equation (1)
are taken to be the solar system r-process abundances
\citep{arl99} for the elements heavier than Ba; for the other
lighter nuclei we use solar system r-process abundances multiplied
by a factor of 0.4266.

\section{Results and Discussion}
Using the observed data in 12 sample stars
\citep{bar05,hil00,coh03,aoki01,aoki02,iva05}, the model
parameters can be obtained. The results of the neutron exposures,
overlap factor, and the component coefficients are listed in table
1.

Figure 1 shows our calculated best-fit results. For most stars ,
the curves produced by the model are consistent with the observed
abundances within the error limits. The agreement of the model
results with the observations provides strong support to the
validity of the parametric model adopted in this work.

In the AGB model, the overlap factor \emph{r} is a fundamental
parameter. The overlap factor deduced for the double-enhanced
stars lies between 0.1 and 0.81. \citet{aoki01} have reported an
overlap factor of r$\sim$0.1 for metal-deficient stars LP 625-44
and LP 706-7; our calculated results of the two stars are close to
their values. For the third dredge-up and the AGB model, several
important properties depend primarily on the core mass
\citep{ibe77,gro93,kar02}. In the core-mass range $0.6\leq M_c
\leq 1.36$, an analytical formula for the AGB stars was given by
\citet{ibe77} showing that the overlap factor increases with
decreasing core mass. Combing the formula and the initial-final
mass relations \citep{zij04}, we can obtain the overlap factor as
a function of the initial mass and metallicity. In AGB stars with
initial mass in the range $M=1.0-4M_\odot$, the core mass $M_c$
lies between 0.6 and 1.4$M_\odot$ at [Fe/H]$\sim$-2.5. According
to the formula by \citet{ibe77}, the corresponding values of
\emph{r} would range between 0.8 and 0.16. In an evolution model
of AGB stars, a small \emph{r} may be realized if the third
dredge-up is deep enough for the s-processed material to be
diluted by extensive admixture of unprocessed material. In fact,
more recent AGB models showing the third dredge-up for low-mass
stars at solar metallicity have found \citep{gal98}, for core
masses around 0.7 M$_\odot$, much lower overlap factors, down to
$r\sim0.4$, than the value of 0.8 given by \citet{ibe77}.
\citet{kar02} and \citet{her00,her04} have found that the third
dredge-up is more efficient for the AGB stars with larger core
masses, confirming the low values of \emph{r} obtained by
\citet{ibe77} in these cases. Taking into account the core-mass
dependence, the wide range of r-values for the double-enhanced
stars that we obtain could possibly be explained by a wide range
of core-mass values of AGB stars at low metallicity.

The neutron exposure per pulse $\Delta\tau$ is also a fundamental
parameter in the AGB model. The neutron exposure per pulse
$\Delta\tau$ deduced for the double-enhanced stars lies between
0.45 and 0.88 mbarn$^{-1}$. \citet{aoki01} have reported a neutron
exposure per pulse $\Delta\tau$ $\sim$ 0.71 mbarn$^{-1}$ for the
metal-deficient star LP 625-44 and $\Delta\tau$ $\sim$ 0.80
mbarn$^{-1}$ for LP 706-7; our calculated results for these two
stars are close to their values. In the case of multiple
subsequent exposures the mean neutron exposure is given by
$\tau_0=-\Delta\tau/$ln$r$. For the s-process nucleosynthesis in
the AGB stars, the final abundance distributions depend mainly on
the mean neutron exposure $\tau_0$. Basically, for a given neutron
exposure one can obtain a higher mean neutron exposure simply by
increasing the overlap factor, because there is more material
subsequently exposed to neutron fluxes, which favors the
production of the heavier elements, such as Pb. This implies that
the formation of a Pb star ([Pb/Ba]$\geq$1.0) could be explained
by a transfer of matter from the low-mass AGB star
(M$<$3.0M$_\odot$). However, one should remember that actually in
the model with the $^{13}$C pocket burning radiatively it is
mostly the amount of $^{13}$C that determines [Pb/Ba]. Introducing
a large range of the $^{13}$C amount at a given metallicity also
proved to be effective for explaining the large scatter of [Pb/hs]
observed in the metal-poor AGB stars \citep{str05}.

The presence of \emph{s}-process elements, along with large
enhancements of carbon, suggests that a mass-transfer episode from
a former AGB star in a binary system took place. Thus, one major
goal is to find an astrophysical scenario, associated with an AGB
star in a binary system, in which the r-process might also occur.
For a binary system with small orbital separation, the matter
accreted from the initial massive star is large \citep{bof88}, so
the level of both s-enrichment and r-enrichment of the observed
star should be high if the r-process also occurs in the binary
system (e.g. AIC or SN 1.5). We compare the strength of the
r-process elements parameterized by C$_r$ with the strength of the
s-process elements parameterized by C$_s$ for the double-enhanced
stars. A surprising result shown in Figure 2 is the strong
correlation between C$_r$ and C$_s$. This implies an increase of
s-process matter accreted from the AGB star with increasing
r-process matter accreted from the AIC or SN 1.5. The correlation
obtained from this work is significant evidence for the
\emph{s}+\emph{r} scenario.

Based on the dependence of the initial-final mass relations on
metallicity, \citet{zij04} speculates that the metal-poor AGB
stars with initial mass in excess of $3-4 M_\odot$ may cause the
degenerate cores to reach the Chandrasekhar mass, leading to
type-1.5 supernovae; this process would explain the enhancement
pattern of the double-enhanced halo stars produced by small
overlap factors (i.e.\ r$\sim0.1-0.2$). Since AIC is thought to
occur only under somewhat restricted conditions, i.e., both the
white dwarf mass and the mass transfer rate must be high enough,
\citet{bai90} assumed that the initial masses leading to heavy
white dwarfs required for AIC would be in the range of
$4M_\odot<M<8M_\odot$ and speculated that the rate of the
occurrence of the AIC process in the Galaxy is no more than
$\sim$10$^{-4}$yr$^{-1}$. This rarity seems to be in conflict with
the substantial fraction of double-enhanced stars ($\sim$30\%)
among all the CEMP$-$\emph{s} stars currently observed.
\citet{coh03} proposed that AIC may be caused in binary systems
with moderately massive AGB stars (M$>$3M$_\odot$) and small
initial orbital separation. Because the core mass of the AGB star
is remarkably large at low metallicity \citep{zij04}, we propose
that the binary systems with lower mass AGB stars
(M$<$3$-$4M$_\odot$) may also cause AIC, which can increase the
rate of the occurrence of AIC and explain the enhancement pattern
of the double-enhanced halo stars produced by large overlap
factors ($r\sim0.3-0.8$). In this case, the orbital separation
must be small enough to allow for capture of a sufficient amount
of material to create the formation of AIC. The initial-final mass
relation flattens at higher metallicity \citep{zij04}, and the
degenerate cores of high-metallicity AGB stars are smaller than
those of the low-metallicity stars, so the formation of AIC or SNe
1.5 is more difficult in high-metallicity binary systems, which
could explain the upper limit of the metallicity ([Fe/H]$<$-2.0)
for the observed double-enhanced stars.

The binary systems with low-mass AGB stars (M$<$3M$_\odot$) and
large initial orbital separation could not cause AIC, because the
white dwarf accretes matter insufficiently from the polluted star
and the companion is polluted only by the former AGB star, which
can explain the formation of \emph{s} stars. For example, the
r-process component coefficients of HD 196944 and CS 30301-015 are
0.6 and 0.9, respectively, which implies that the two stars are
\emph{s} stars. The s-process component coefficients of these
stars are smaller than those of the other 10 stars. This fact
illustrates that there is less accreted material of these stars
from the former AGB stars, and they should be formed in the binary
systems with larger initial orbital separations. In addition, the
\emph{s}-element patterns of these two stars are matched by models
with r=0.44 and 0.34, which are large enough to be consistent with
the idea that the AGB star polluting these \emph{s} stars was of
low mass.

The \emph{r} star could possibly be formed in a close binary
system, favoring an early mass transfer from the massive companion
in its RGB(red giant branch) phase. In this case, the massive
companion should produce no \emph{s}-elements, and its final
explosion produces \emph{r}-elements \citep{bar05}. The \emph{r}
star could be also formed in a molecular cloud that was polluted
by a single supernova or a few supernovae \citep{shi98,fie02}. The
halo star with perhaps the most striking r-process composition is
CS 22982-052; its r-process abundance ratios are
\begin{equation}
 [\frac{r}{Fe}]\simeq1.7\Rightarrow \frac{r}{Fe}\simeq50(\frac{r}{Fe})_{\odot} .
\end{equation}
This indicates C$_{r}$$\approx$50, in the range of component
coefficients C$_r$ obtained for the double-enhanced stars.

As an example, it is interesting to investigate a possible
explanation of the parameters obtained for a double-enhanced star
using the s+r scenario. The s-process abundances of HE 2148-1247
are a result of pollution from the dredged-up material of the
former AGB star. The measured [\emph{s}/Fe] refers to the average
s-processed material of the AGB star after dilution by mixing with
the envelope of the presumed unevolved companion that is now the
extrinsic star. The parameters deduced for HE 2148-1247 are r=0.1,
$\Delta\tau$=0.88 mbarn$^{-1}$, and $C_s=0.0045$. Adopting the
analytical formula given by \citet{ibe77} and the initial-final
mass relations \citep{zij04}, we can know that the initial mass of
the former AGB star is about 3$-$4 M$_\odot$, which lies in the
range of $M\sim3-12M_\odot$ reported by \citet{coh03}. At some
time on the AGB, the convective He-shell and the envelope of the
giant will be overabundant in heavy elements by factors
$f_{shell}$ and $f_{env,\ 1}$, respectively, with respect to solar
system abundances normalized to the value of [Fe/H]. The
approximate relation between $f_{env,\ 1}$ and $f_{shell}$ is
\begin{equation}
f_{env,\ 1}\approx\frac{\Delta M_{dr}}{M^e_1}f_{shell} ,
\end{equation}
where $\Delta$$M_{dr}$ is the total mass dredged up from the He
shell into the envelope of the AGB star and $M^{e}_{1}$ is the
envelope mass of the AGB star. For a given s-process element, the
overabundance factor $f_{env,\ 2}$ in the future \emph{s}+\emph{r}
star envelope can be approximately related to the overabundance
factor $f_{env,\ 1}$ by
\begin{equation}
 f_{env,\ 2}\approx\frac{\Delta M_{2}}{M^e_2}f_{env,\ 1}\approx\frac{\Delta M_2}{M^e_2}\frac{\Delta M_{dr}}{M^e_1}f_{shell} ,
\end{equation}
where $\Delta$$M_{2}$ is the amount of matter accreted by the
future \emph{s}+\emph{r} star and $M^{e}_{2}$ is the envelope mass
of the star. The component coefficient $C_{s}$ is computed from
the relation
\begin{equation}
 C_s =\frac{f_{env,\ 2}}{f_{shell}}\approx\frac{\Delta M_2}{M^e_2}\frac{\Delta
 M_{dr}}{M^e_1} .
\end{equation}
The value of $f_{shell}$ of course depends on the heavy element
considered. For Pb, the factor $f_{shell}$ of HE 2148-1247 derived
in the present study is about 1.6$\times$10$^{5}$, which lies in
the $6.3\times10^{4} \leq f_{shell} \leq 2.5\times10^{5}$ range
for the metal-poor AGB stars with metallicities from [Fe/H]=-2.0
down to -3.0 calculated by \citet{bus01}. For the sake of
simplicity, no continuous variation of the factor during the AGB
phase has been considered. Taking $\Delta M_{dr}=0.01M_{\odot}$
\citep{bus92}, $\Delta\dot{M}_{2}\approx-0.2\dot{M}_1$
\citep{ste05}, and $M^{e}_{2}\approx0.5M_{\odot}$ \citep{bof88},
the value of $C_{s}=0.004$ is deduced from equation (5), which is
close to the value for HE 2148-1247. Because of the uncertainties
of mass-loss rates and our poor knowledge of how and when mass
transfer phenomena occur, we do not claim that this is the only or
even the best understanding of the parameters.

We discuss the uncertainty of the parameters using the method
presented by \citet{aoki01}. Figure 3 and 4 show the calculated
ratios log(Pb/Ba) and log(Ba/ls), where \emph{ls} denotes the
`light' s-process elements (Sr and Y), as a function of the
neutron exposure $\Delta\tau$ in a model with r=0.1 and a function
of overlap factor \emph{r} with a fixed neutron exposure
$\Delta\tau$=0.88 mbarn$^{-1}$. These are compared with the
observed ratios from HE 2148-1247. There is only a region of
overlap in Figure 3, $\Delta\tau$ =0.88$\pm$0.08 (1 $\sigma$)
mbarn$^{-1}$, in which both the observed ratios log(Pb/Ba) and
log(Ba/ls) can be accounted for. The bottom panel in Figure 3
displays the reduced $\chi^2$ value calculated in our model and
there is a minimum, with $\chi^2 =2.36$, at $\Delta\tau =0.88
mbarn^{-1}$ with a 1 $\sigma$ error bar $\pm$0.08 mbarn$^{-1}$;
the neutron exposure is constrained quite well. The ratios
log(Pb/Ba) and log(Ba/ls) are insensitive to the overlap factor
\emph{r} and allow for a wider range, $0<r<0.28$. The
uncertainties of the parameters for the star HE 2148-1247 are
similar to those for metal-poor stars LP 625-44 and LP 706-7
obtained by \citet{aoki01}. For most sample stars, the
uncertainties of the parameters are smaller than those for HE
2148-1247.

Our model is based on the observed abundances of the
double-enriched stars and nucleosynthesis calculations, so the
uncertainties of those observations and measurement of the
neutron-capture cross sections will be involved in the model
calculations. We note from table 1 that for three stars (HE
2148-1247, LP 625-44, and CS 29497-030), the reduced $\chi^2$ are
about 2. The probability that $\chi^2$ could be this large as a
result of random errors in the measurement of the neutron-capture
cross sections and the abundances of the neutron-capture elements
is less than 2\%. We find that all these uncertainties cannot
explain the larger errors of neutron-capture elements, such as Zr
in HE 2148-1247 and Y in CS 29497-030. This implies that the
understanding of the true nature of the s-process or r-process is
incomplete for at least some of these elements \citep{tra04}.

\section{ Conclusions}
 A main result  of this work is the wide range of
r-values for the double-enhanced stars. This can be explained when
varying the initial mass of the AGB stars. We also find a strong
correlation between the strength of the r-process elements and the
strength of the s-process elements for the double-enhanced stars.
This correlation is significant evidence for the scenario in which
both the r- and s-enhancements of \emph{s}+\emph{r} stars are
produced in the binary system. Since the formation of the AIC or
SN 1.5 is more difficult for a binary system with high
metallicity, the upper limit of the metallicity ([Fe/H]$<$-2.0)
for the observed double-enhanced stars could be explained.

Based on the discussion above, the double-enhanced stars belong to
binaries. If these stars are found to be nonbinaries, they may
also have been binaries initially, ceasing to be binaries due to
the explosion of the massive companion. Overall, the number of
CEMP stars is still small, especially for the CEMP-\emph{r} stars,
making proposed enrichment scenarios difficult to explicitly test.
To underpin these studies, accurate abundance analyses for similar
s- and r-process-rich, metal-poor, carbon-enhanced stars are
required. Obviously, a more precise overlap factor-core mass law
and initial-final mass relations from detailed models of AGB stars
are needed. Further abundance studies of neutron-capture-rich
stars will reveal the characteristics of the s- and r-processes at
low metallicity, such as their metallicity and mass dependence,
and the history of enrichment of neutron-capture elements in the
early Galaxy.

\acknowledgments

We are grateful to the referee for very valuable comments and
suggestions that improved this paper greatly. This work has been
supported by the National Natural Science Foundation of China
under grant 10373005.

\clearpage


\begin{figure}
\epsscale{1.00} \plotone{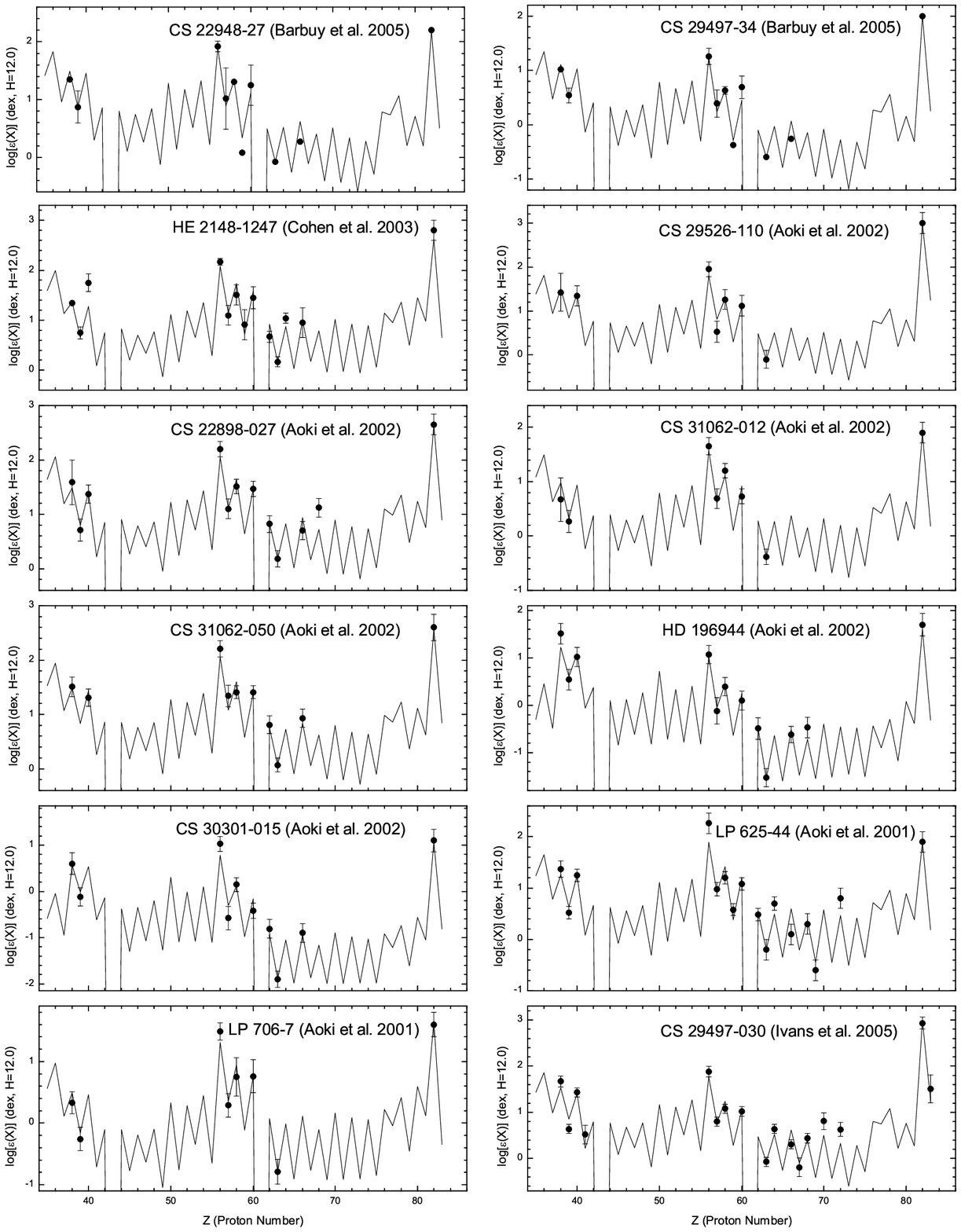} \caption{{Best fit to
observational results of metal-deficient stars. The black circles
with appropriate error bars denote the observed element
abundances, and the solid lines represent predictions from
s-process calculations considering r-process
contribution.}\label{fig1}}
\end{figure}

\begin{figure}
\epsscale{.80} \plotone{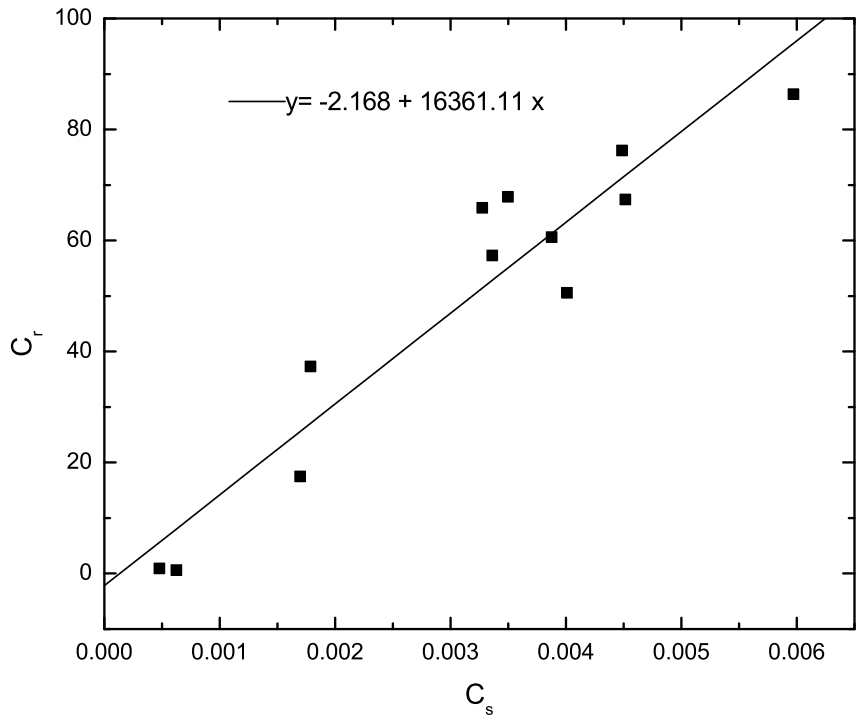} \caption{{Correlation between the
s-process component coefficients C$_s$ and the r-process component
coefficients C$_r$ for the double-enhanced stars. The fit to the
relation is shown as a solid line. }\label{fig2}}
\end{figure}

\begin{figure}
\epsscale{.80} \plotone{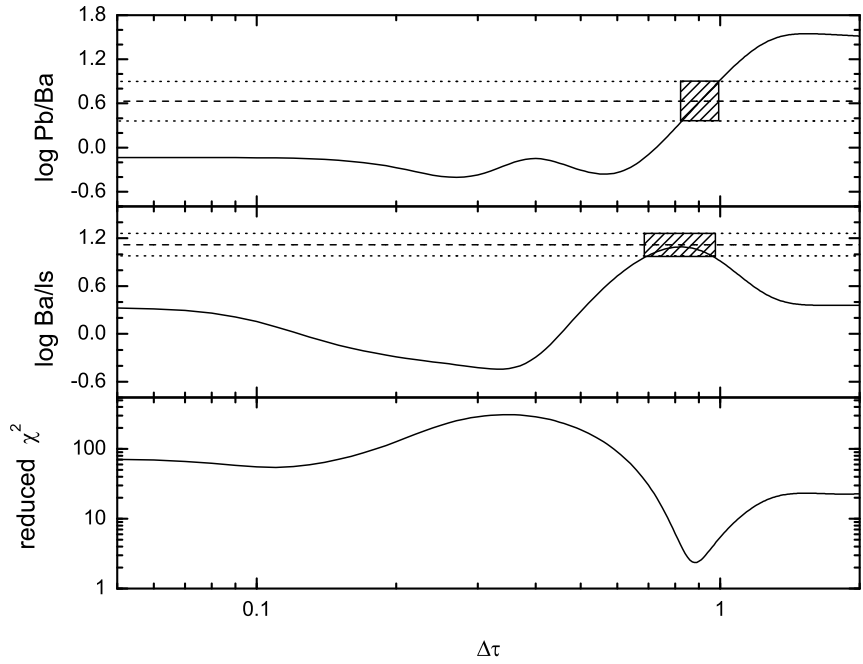} \caption{{Abundance ratios log
(Pb/Ba) (\emph{top}), log (Ba/ls) (\emph{middle}), and reduced
$\chi^2$ (\emph{bottom}), as a function of the neutron exposure
per pulse, $\Delta\tau$, in a model with overlap factor $r=0.1$.
Solid curves refer to the theoretical results, and dashed
horizontal lines refer to the observational results with errors
expressed by dotted lines. The shaded area illustrates the allowed
region for the theoretical model. }\label{fig3}}
\end{figure}

\begin{figure}
\epsscale{.80} \plotone{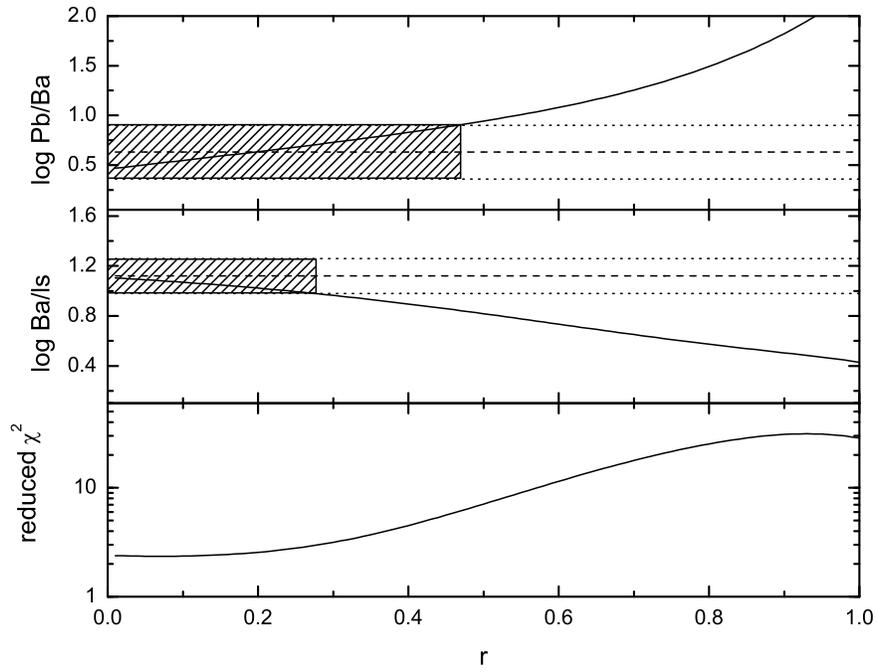} \caption{{Same panels as those in
Fig.\ 3, but as a function of the overlap factor \emph{r}.
}\label{fig4}}
\end{figure}

\clearpage

\begin{deluxetable}{lccccccclrc}
\tabletypesize{\footnotesize}
 \tablecolumns{13} \tablewidth{0pc}
\tablecaption{Observed abundance ratios and the derived parameters
for s- and r-rich stars} \tablehead{ Star & [Fe/H] &[Ba/Fe]
&[Eu/Fe] &[Pb/Ba] &\emph{r} &$\Delta\tau $ &$\tau_0$ &$C_s$ &$C_r$
&$\chi^2$\\
&  & & & & &(mbarn$^{-1}$) &(mbarn$^{-1}$) & & & }
 \startdata
 CS 22948-27 &-2.47  &2.26 &1.88
&0.46  &0.37 &0.61  &0.614  &0.0033  &65.9 &0.402859\\
 CS 29497-34  &-2.90 &2.03 &1.80  &0.92  &0.61
&0.53  &1.072  &0.0034  &57.3 &1.100821\\
 HE 2148-1247  &-2.30  &2.36 &1.98  &0.76
&0.10 &0.88  &0.382  &0.0045  &67.4 &2.360219\\
 CS 29526-110  &-2.38 &2.11 &1.73  &1.19
&0.79 &0.64  &2.715  &0.0040 &50.6 &0.580903\\
 CS 22898-027  &-2.25 &2.23 &1.88  &0.61
&0.42 &0.77 &0.888  &0.0035  &67.9 &1.225389\\
 CS 31062-012  &-2.55 &1.98 &1.62  &0.42
&0.32 &0.71 &0.623  &0.0018  &37.3 &0.883710\\
 CS 31062-050  &-2.32 &2.30 &1.84  &0.60
&0.45 &0.71 &0.889  &0.0039 &60.6 &1.027867\\
 HD 196944  &-2.25  &1.10 &0.17  &0.80
&0.44 &0.45 &0.548  &0.0006 &0.6 &0.586861\\
 CS 30301-015  &-2.64 &1.45 &0.2  &0.25
&0.34 &0.54 &0.501  &0.0005 &0.9 &1.118641\\
 LP 625-44  &-2.71  &2.74 &1.97  &-0.19
&0.16 &0.69  &0.377  &0.0045 &76.2 &2.111419\\
 LP 706-7  &-2.74  &2.01 &1.40  &0.27
&0.10 &0.82  &0.356  &0.0017 &17.5 &0.846228\\
CS 29497-030  &-2.57 &2.32 &1.99  &1.33
&0.81 &0.61  &2.895 &0.0060 &86.4 &2.441984\\
\enddata
\end{deluxetable}

\clearpage

\end{document}